\documentclass[prb,aps,superbib,tightenlines, twocolumn,floatfix,superscriptaddress]{revtex4}
\usepackage{natbib}
\usepackage{graphicx}
\begin{document}

\title{Electronic Excitations and Insulator-Metal Transition in Poly(3-hexylthiophene) Organic Field-Effect Transistors}
\author{N. Sai}
\affiliation{Department of Physics, University of California, San
Diego, La Jolla, CA 92093}
\author{Z.Q. Li}
\affiliation{Department of Physics, University of California, San
Diego, La Jolla, CA 92093}
\author{M.C. Martin}
\affiliation{Advanced Light Source Division, Lawrence Berkeley
National Laboratory, Berkeley, California 94720}
\author{D.N. Basov}
\affiliation{Department of Physics, University of California, San
Diego, La Jolla, CA 92093}
\author{M. Di Ventra}
\affiliation{Department of Physics, University of California, San
Diego, La Jolla, CA 92093}
\date{November 9, 2006}
\pacs{72.80.Le, 78.30.-j, 71.30.+h, 71.38.-k}

\begin{abstract}
We carry out a comprehensive theoretical and experimental study of charge
injection in Poly(3-hexylthiophene) (P3HT) to determine the most likely scenario
for metal-insulator transition in this system. We calculate the
optical absorption frequencies corresponding to a polaron and a bipolaron lattice
in P3HT. We also analyze the electronic excitations for three possible scenarios under which
a first-- or a second--order metal--insulator transition can occur in doped
P3HT. These theoretical scenarios are compared with data from infrared
absorption spectroscopy on P3HT thin film field-effect transistors (FET).  Our
measurements and theoretical predictions suggest that charge-induced localized
states in P3HT FETs are bipolarons and that the highest doping level achieved in our
experiments approaches that required for a first-order metal--insulator
transition.
\end{abstract} \maketitle

\section{Introduction}

\begin{figure}
\includegraphics[width=5cm]{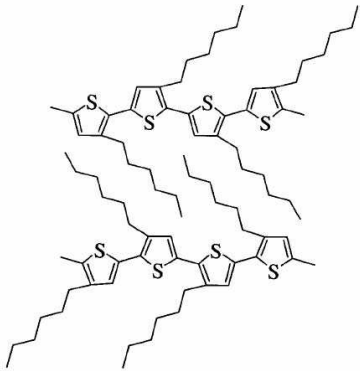}
\caption{Schematic diagram of RR P3HT.}
\label{fig:structure}
\end{figure}

Upon doping the conductivity of conjugated organic polymers can be controlled over an
extraordinarily large range from insulator to metal.\cite{Kohlman98p85} Therefore, these
materials hold great promise for high--performance organic
electronics.\cite{Dimitrakopoulos02p99,Forrest05p911,Malliaras05p53} The organic field-effect
transistor is a benchmark system for exploring the fundamental properties of the organic
semiconductors as well as for exploiting their novel functionalities.\cite{Horovitz98p365}
Recently, particular interest has been devoted to FETs based on highly regio-regular (RR)
P3HT (see Fig.~\ref{fig:structure}).\cite{Bao96p4108,Sirringhaus98p1741,Sirringhaus99p685,Osterbacka00p839,Wang04p316,Taniguchi04p3298,Hamadani05p235302,Street05p165202,Kline06p222,Panzer06p1051,Dhoot06p246403}
The charge carrier mobility in these devices has been found as high as $0.7$ cm$^2$/Vs, enhanced
by orders of magnitude compared to that of the more disordered P3HT films. There has also been a
strong interest in using the voltage in the organic FETs to induce a sufficiently high density
of carriers so that metallic states can be produced in the
polymers.\cite{Panzer06p1051,Dhoot06p246403}

Unlike traditional semiconductors, charge injection into organic polymers induces structural
deformations in the polymer chains and gives rise to self-localized excitation states inside the
band gap.  The Peierls instability in {\it trans}--polyacetylene (PA) leads to
solitons,\cite{Heeger88p781} or kink-like excitations with a two--fold-degenerate ground state.
However, most of the conjugated polymers are analogous to {\it cis}--PA, which consists of two
non-degenerate ground states (NDGS); i.e., the two conformations resulting from the mutation of
a single and a double C--C bond are not energetically equivalent. Two possible excitations in
NDGS conjugated polymers are polarons or bipolarons that are associated to a pair of bound
states in the band gap split off from the continuum.\cite{Heeger88p781} Experimental evidence of
similar signatures have been reported in P3HT and have been interpreted in terms of both types
of excitations. Earlier work on chemical doping and photoinduced absorption in P3HT have
assigned the two localized intragap states observed to bipolarons,\cite{Kim88p5490} whereas a
more recent study of RR-Poly(3-alkylthiophene) concluded that while chemical doping in P3HT
films results in localized polarons, photoexcitations give rise to delocalized
polarons.\cite{Osterbacka00p839} A charge modulation spectroscopy experiment in regio-random
P3HT in a metal--insulator--semiconductor diode has identified five sub-gap transition features,
out of which two are associated to optical transitions of doubly-charged bipolarons and three to
singly-charged polarons, suggesting coexistence of the two charge species.\cite{Ziemelis91p2231}
More recently, however, the same technique applied to RR P3HT identified only three sub-gap
features, which were assigned to polarons.\cite{Sirringhaus99p685,Brown01p125204}

Metallic states attested by a finite density of states at the Fermi energy have been reported in
chemically-doped P3HT polymers using photoemission spectroscopy and have been attributed to the
polaron lattice.\cite{Logdlund89p1841} On the other hand, there exists a model which supports
the bipolaron lattice in NDGS polymers, arguing that the metallic states are a consequence of
the merging of the bipolaron band with either the valence or the conduction
band.\cite{Conwell91p937} However, in neither of these studies a prediction of the critical
doping concentration under which P3HT becomes metallic has been given.

Here we report a joint theoretical and experimental study of charge-induced electronic
excitations in P3HT. We present models for three scenarios under which charge-induced
transitions between the bipolaron lattice and the polaron lattice or the merging between the
localized states and the extended states can take place. We predict the critical doping
thresholds above which metallic states can be expected in charge-injected P3HT. To this end, we
apply the continuum Hamiltonian model introduced by Brazovskii an Kirova\cite{Brazovskii81p4}
(BK) for that takes into account the electronic and lattice couplings. We carried out infrared
spectroscopic studies in P3HT- based FETs. From these measurements we extract the
charge-injected doping levels and compare with the predicted critical values, which provide a
lower bound for metallic behavior. The analysis suggests that the charge carriers in our P3HT
FET are bipolarons and the highest charge injection level reaches the proximity of the critical
doping value that is needed for metallic states in P3HT. We also discuss our results in view of
very recent experimental evidence of field--induced metallic behavior in P3HT
FETs,\cite{Panzer06p1051} and a structurally similar polymer with polyelectrolyte gate
dielectric.\cite{Dhoot06-2}

The paper is organized as follows. In Sec.~\ref{model}, we briefly review the solution of the BK
model as derived in the literature and provide an analytical formulation for bandwidths of
polarons and bipolarons. In Sec.~\ref{theory}, the solutions are applied to three scenarios
involving polaron and bipolaron states in P3HT. From this study the critical dopings required to
achieve metallic states in P3HT are predicted. In Sec.~\ref{exp}, we present experimental
results on charge injection in P3HT-based transistors and we compare them with the theoretical
predictions in Sec.~\ref{dis}.

\section{Theoretical Formulation}
\label{model}
\subsection{Polarons and bipolarons}

The Su-Schrieffer-Heeger (SSH)\cite{Su79p1698} electron-phonon coupling model and its continuum
version\cite{Takayama80p2388} have been widely used to study structural excitations (solitons,
polarons) in degenerate polymers such as {\it trans}--PA. In NDGS polymers, however, the
degeneracy is broken. A generalization of the SSH model by Brazovskii and Kirova has been
proposed to take into account the one-electron energy gap in the absence of dimerization. The
solution of this model has been derived and discussed
elsewhere;\cite{Brazovskii81p4,Fesser83p4804,Onodera84p775} here we briefly summarize the main
results.

The gap parameter contains both an intrinsic and an extrinsic contribution $\Delta(x) = \Delta_e
+ \Delta_i(x)$, where $\Delta_i(x) $ is associated with the dimerization due to the
$\pi$-electron- lattice interaction and $\Delta_e$ is a constant gap parameter that arises from
the $\sigma$-bond orbitals. The BK continuum Hamiltonian is written as \begin{eqnarray} H &=&
\int dx \frac{\Delta_i(x)^2}{\pi\lambda v_F} \nonumber\\ &+&\int dx \Psi^\dagger(x)\left[-iv_F
\sigma_3 \frac{\partial} {\partial x} + \Delta(x)\sigma_1\right] \Psi(x) . \label{eq:Ham}
\end{eqnarray} Here, the first term describes the interaction of the electrons with the lattice
deformation $\Delta_i(x)$, $\Psi(x)$ is the two-components fermion field, $\lambda$ is the
dimensionless electron-phonon coupling strength, and $v_F$ is the Fermi velocity of the $\pi
$-electrons. The second term describes the self-energy of the $\pi$-electrons in a field of a
dimerization potential $\Delta(x)$, while $\sigma_i$ are the Pauli matrices.

In the continuum limit, the analytical solution of the BK Hamiltonian
represents a polaron or a bipolaron,
\begin{equation}
\Delta(x) = \Delta_0 + k v_F \{ \tanh[k(x-x_0)] - \tanh[k (x+x_0)]\} ,
\label{eq:gap}
\end{equation}
with
\begin{equation}
\tanh(2k x_0) = \frac{kv_F}{\Delta_0}
\end{equation}
and
\begin{equation}
\Delta_0 =  W e^{-1/\lambda} e^\gamma ,
\end{equation}
where $W$ is the width of the $\pi$-band and $\gamma=\frac{\Delta_e}
{\lambda \Delta_0}$ is the confinement parameter that describes the
extent of the energy degeneracy breaking. 

There are two bound states corresponding to the polaron solution in
Eq.~(\ref{eq:gap}) with eigenvalues symmetrically placed in the gap
at $\pm\omega_0$ where
\begin{equation}
\omega_0 = \Delta_0\sqrt{1-(k\xi_0)^2}
\label{eq:omega0}
\end{equation}
and $\xi_0=\frac{ v_F}{\Delta_0}$ is the parameter that describes the spatial extent of the soliton kink. Minimizing the energy of
Eq.~(\ref{eq:Ham}) with respect to $\omega_0$ yields the following self-consistent condition for
$k$:
\begin{equation}
\frac{kv_F}{\omega_0}\gamma = \sin^{-1} (\frac{\omega_0}{\Delta_0}) +
\frac{\pi}{4} (n_+ - n_-),
\label{eq:selfcon}
\end{equation}
where $n_\pm$ are the occupation numbers of the discrete energy
levels. For a singly-charged system such as an electron polaron or a
hole polaron, $n_+ - n_- = -1$, and for a doubly charged system such
as a bipolaron, $n_+ - n_- = 0$.
The polaron (bipolaron) formation energy can be obtained from the
total energy given in the above configuration\cite{Onodera84p775}
\begin{eqnarray}
E_{p,b} &=& \frac{4}{\pi} \left[(1-\gamma) kv_F + 2\gamma\Delta_0kx_0+
\omega_0\sin^{-1}\left(\frac{\omega_0}{\Delta_0}\right)\right]
\nonumber\\
&&+ (n_+ - n_-)\omega_0.
\label{eq:Ep}
\end{eqnarray}

\subsection{Bandwidth of polarons and bipolarons}

As the doping level increases, the charges stored on individual polaron states overlap and give
rise to the polaron band. The bandwidth can be calculated from the off-diagonal element of the
BK Hamiltonian between unperturbed states on each polaron.\cite{Kivelson85p308} Here we follow
the approach of Onodera\cite{Onodera84p775} in which the interaction between two hole-doped
polarons approaching each other are considered. When the doping level increases, the overlap of
electron wave functions trapped in the two polaron dips in $\Delta(x) $ gets stronger and the
two bound state energy levels $\pm\omega_0$ are each split into two levels $\pm\omega_1$ and
$\pm\omega_2$ (bonding and anti--bonding states). The system can lower its energy by
accommodating both holes on the lowest level.

In a tight--binding approximation, the bandwidth of the polarons can
then be calculated from the difference between the two discrete
levels. The gap parameter for the two polaron or two bipolarons is
found to be\cite{Onodera84p775}
\begin{eqnarray}
\Delta(x) &=&\Delta_0 + \frac{(k_1^2 - k_2^2) v_F }{k_1\coth(k_1x -
\beta_1) - k_2 \tanh(k_2x - \beta_2)} \nonumber\\
&-& \frac{(k_1^2 - k_2^2)v_F}{k_1\coth(k_1x + \beta) - k_2\tanh (k_2x
+ \beta_2)},
\end{eqnarray}
where
\begin{equation}
\tanh 2\beta_j = k_j\xi_0 (j= 1,2),
\end{equation}
and $k_1$ and $k_2$ are related to the discrete energy levels by
\begin{equation}
\omega_j = \Delta_0 \sqrt{1-(k_j\xi_0)^2}.
\label{eq:omega_j}
\end{equation}

As the two polarons approach each other, $k_1-k_2$ becomes nonzero
and is asymptotically related to the separation $d$ between the two
polarons via
\begin{equation}
k_1-k_2 = 4k e^{-dk}
\end{equation}
at large distances.  Therefore the separation between the bonding and
antibonding states can be calculated from Eq.~(\ref{eq:omega_j})
which now becomes
\begin{equation}
(k1^2-k_2^2)v_F^2 + (\omega_1^2-\omega_2^2) = 0.
\end{equation}
If the separation $d$ is large, $k_1 - k_2$ is small, we can rewrite
this relation as
\begin{equation}
k(k_1-k_2) v_F^2 + \omega_0(\omega_1 - \omega_2) = 0.
\end{equation}
Replacing $d$ by the dopant concentration $y = a/d$, where $a$ is the
projected bond length along the chain axis, we obtain the bandwidth
of the polarons
\begin{eqnarray}
W_p(y) = -4(\omega_0 - \omega_2)\simeq  8\frac{k^2v_F^2}{\omega_0} e^
{-ka/y}.
\label{eq:Wp}
\end{eqnarray}
In a similar fashion, the two-bipolaron system has been
solved,\cite{Onodera84p775} and an asymptotic expression of the bipolaron
bandwidth similar to that for the polarons can be obtained,
\begin{equation}
W_{b}(y) \simeq \frac{32}{\pi} \left( 1+ \frac{\gamma\Delta_0^2}
{\omega_0^2}\right) \frac{k^3v_F^3 }{\omega_0^2} e^{-2ka/y}.
\label{eq:Wbp}
\end{equation}
Expressions ~(\ref{eq:Wp}) and ~(\ref{eq:Wbp}) represent our starting point
for the discussion that follows.

\section{Theoretical Results}
\label{theory}

\begin{figure}
\includegraphics[width=9cm]{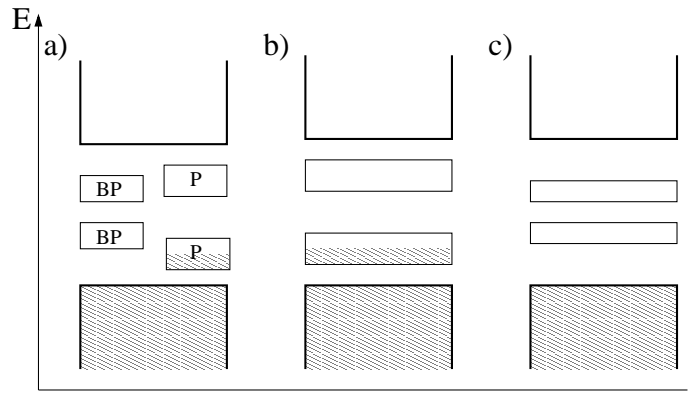}
\caption{Schematic diagram of the localized gap states in a P3HT polymer with p-type doping
showing three scenarios for the metal-insulator transition including: a) the crossover between
the bipolaron and the polaron formation energy (the lower polaron band is partially filled) upon doping; b) the polaron bands broaden to merge with the continuum; c) the bipolaron band broaden to merge with the continuum. The shades are indicative of occupied electronic bands.}
\label{fig:band}
\end{figure}

In NDGS polymers, doped or injected charges can lead to the formation of either polarons or
bipolarons, depending on the formation energy of the excitations and the relative energy
difference between the two configurations upon doping. Here we apply the solutions to the BK
model to predict the effect of the doping concentration on the electronic excitations in P3HT.
We considered three types of band diagrams as shown in Fig.~\ref{fig:band}. They correspond to
a) a first-order transition between a bipolaron and a polaron lattice, b) a gap closure of the
polaron band and the conduction or the valence band, and c) a gap closure between the bipolaron band
and the conduction or the valence band. In the latter two cases, we look for the condition that the
polaron (bipolaron) bands merge with the extended states and we expect these transitions to be
of second order.

\subsection{Optical absorption energies}
\label{energy}
Before proceeding to calculate the critical doping concentration that
can give rise to metallic states, we calculate the localized energy
levels and the formation energy for the polarons and bipolarons of
P3HT polymers. The two excitation states correspond to a pair of
symmetrically localized levels\footnote{For P3HT, the two localized levels are not
exactly symmetric due to the Sulfur atom that breaks the electron-hole
symmetry, but the difference should be small.} in the gap that
can be calculated from Eq.~(\ref{eq:omega0}). To solve for $k$, we
apply the self-consistency condition Eq.~(\ref{eq:selfcon}),
\begin{equation}
\frac{\gamma z_p}{\sqrt{1-z_p^2}} = \sin^{-1} \sqrt{1-z_p^2} - \frac
{\pi}{4}
\label{eq:selfconf_p}
\end{equation}
and
\begin{equation}
\frac{\gamma z_b}{\sqrt{1-z_b^2}}= \sin^{-1} \sqrt{1-z_b^2},
\label{eq:selfconf_b}
\end{equation}
for the polaron and the bipolaron configuration, respectively, where $z= k\xi_0$.
The confinement parameter $\gamma$ can be either estimated from the energy
splitting between the doping-induced gap states\cite{Fesser83p4804} or by solving
for the phonon propagator from Raman resonance dispersion
data.\cite{Horovitz82p729} In either method, $\gamma$ is estimated to be $0.3\sim
0.4$ in P3HT (we discuss later the effect of disorder on this
parameter).\cite{Kim88p5490} Solving the above two equations, we find that $z_p$
is in the range $0.553-0.517$ and $z_b$ in the range $0.867-0.829$.

The energy gap of P3HT has been determined to be $2\Delta_0=2.1$ eV.\cite{Kim88p5490} Together
with the above values of $z$, we find the localized energy levels $\omega_0$ that correspond to
the polaron and the bipolaron. The singly-occupied polaron states allow three optical absorption
transitions, namely, a transition between the valence band and the $-\omega_0$ level
($\omega_1$), a transition between the valence band and the $+\omega_0$ level ($\omega_2$) and a
transition between the two localized levels ($\omega_3$).\cite{Fesser83p4804} For the bipolarons
instead, the last type of transition between the two localized levels is forbidden because the
localized states are either both empty or doubly occupied.  Using the value of $\gamma=0.4$ in
P3HT, we calculated the optical transition frequencies corresponding to the three transitions
for the polaron and the two allowed transitions for the bipolarons in Table~\ref{tab:energy}.

Substituting the self-consistent condition into Eq.~(\ref{eq:Ep}), we
also find that the formation energy of a polaron and a bipolaron can
be expressed as $E_{p,d}=\frac{4}{\pi}\Delta_0N(z)$ with $N(z) = z+
\gamma \tanh^{-1}(z)$ and their values are shown in Table~\ref{tab:energy}.
\begin{table}[tbp]
\caption{The Eigenvalues, the subgap optical transition energies and the formation energies of a polaron and a bipolaron in P3HT calculated from the BK model. The values in the parenthesis are calculated using $\gamma=0.3$. The units are eV.}
\begin{ruledtabular}
\begin{tabular}{c|ccccc}
excitations & $\omega_0$ & $\omega_1$ & $\omega_2$ & $\omega_3$ & $E$ \\\hline
Polaron     & 0.89 (0.87) & 0.16& 1.94 &1.78 &  1.0\\
Bipolaron   &  0.59 (0.52) & 0.46& 1.74 & - & 1.64 \\
\end{tabular}
\end{ruledtabular}
\label{tab:energy}
\end{table}

\subsection{Transition between bipolarons and polarons}
\label{cross}
We here discuss the transition between the bipolaron lattice and the polaron
lattice (see Fig.~\ref{fig:band}a). This first-order phase transition in NDGS
conjugated polymers has been previously proposed without, however, an explicit
calculation of the critical doping density.\cite{Kivelson85p308,Saxena87p3914}
This transition is similar to the energy crossover between the soliton lattice
and the metallic polaron lattice, as proposed for {\it trans}--PA by Kivelson and
Heeger.\cite{Kivelson85p308} The formation energy of a bipolaron is smaller than
the formation energy of two single polarons, as also verified by the formation
energies that we have calculated in the preceding subsection. Therefore, upon
doping, the charges will be stored in bipolarons rather than the polarons.
However, due to the partial filling, the energy per polaron changes with doping,
whereas the bipolaron energies remain unchanged because the bipolaron bands are
either doublely occupied or empty. The condition to be satisfied by this
transition is
\begin{equation}
W_p(y) = \pi (2 E_p - E_b) ,
\label{eq:transition}
\end{equation}
where $W_p(y)$ is the bandwidth of the polarons we have derived in
Section~\ref{model}. Substituting the formation energy of a polaron
and a bipolaron in the above equation, we get
\begin{eqnarray}
W_p(y) = 4\Delta_0[2N(z_p) - N(z_b)].
\end{eqnarray}
Using the polaron bandwidth, Eq.~(\ref{eq:Wp}), we can solve for the
critical dopant concentration
\begin{eqnarray}
y_c= -\frac{z_p a}{\xi_0}\left\{\ln\frac{\sqrt{1-z_p^2}[2N(z_p) - N(z_b)]}{2z_p^2}\right\}^{-1},
\end{eqnarray}
where $\xi_0$ is the characteristic length that can be
determined from $\xi_0\approx \frac{W}{2\Delta_0} a$,\cite{Kivelson85p308} with
$a$ the lattice spacing and $W$ the $\pi$ band width. Since the $\pi$ bandwidth
is a rather universal constant for most of conjugated polymers, we have taken a
typical value $W\sim10$ eV\cite{Hutchison03p035204} and obtained $\xi_0\approx
4.7a$. Therefore we find that $y_c\sim 0.092$.

\subsection{Polaron gap closure}
\label{pol}
A second possible scenario corresponds to the case in which the
polaron energy levels stay unchanged upon doping but the bandwidth of
the polaron states broadens (see Fig.~\ref{fig:band}b). When the
doping is sufficiently large, the polaron band can merge with the
extended states. This situation can be realized as long as the condition
\begin{equation}
W_p(y) = 2 \omega_1
\label{eq:pol}
\end{equation}
is satisfied. Substituting this in the
expression for the polaron bandwidth we obtain
\begin{equation}
\frac{8z_p^2\Delta_0}{\sqrt{1-z_p^2}} e^{-ka/y} = 2\omega_1
\end{equation}
and thus
\begin{equation}
y_c = -\frac{z_pa}{\xi_0}\left(\ln\frac{\omega_1\sqrt{1-z_p^2}}{4
\Delta_0 z_p^2}\right)^{-1}.
\end{equation}
Using the lowest absorption energy of a polaron state found in Sec.~\ref{energy}, i.e.,
$\omega_1 = 0.16$ eV, we find that the critical doping concentration is $y_c = 0.051$.

\subsection {Bipolaron gap closure}
\label{bip}
Similar to the case of the polaron, merging between the bipolaron
band and the valence band (see Fig.~\ref{fig:band}c) can also occur
if the condition
\begin{equation}
W_{b}(y) = 2\omega_1
\label{eq:bp}
\end{equation}
is satisfied. From the bipolaron bandwidth Eq.~(\ref{eq:Wbp}), the above condition becomes
\begin{equation}
\frac{32}{\pi} \frac{\Delta_0z_b^3 }{1-z_b^2}\left(1+\frac{\gamma}{1-
z_b^2}\right) e^{-2ka/y} = 2\omega_1,
\end{equation}
hence
\begin{equation}
y_c = -\frac{2z_ba}{\xi_0}\left\{\ln\frac{\pi\omega_1 (1-z_b^2)^2}{16
\Delta_0 z_b^3 (1-z_b^2 + \gamma)}\right\}^{-1}.
\end{equation}
Using the result $\omega_1 = 0.46$ eV and the value of $z_b$ for a bipolaron in
Sec.~\ref{energy}, we find that the critical dopant concentration for this transition is $y_c =
0.091$.

\section{Infrared  spectroscopic results }
\label{exp}

In order to study the insulator-metal transition in conjugated polymers in functional organic
FET devices, we carried out systematic investigations of P3HT based FETs employing Infrared (IR)
spectroscopy. The critical advantage of IR spectroscopy is that it directly probes the
electronic excitations associated with the injected carriers in organic FET
devices.\cite{Li06p224} Combined with theoretical analysis in Sec.~\ref{theory}, these
experiments uncover the nature of the electronic excitations in P3HT. Moreover, the carrier
density in the accumulation layer of P3HT can be evaluated from the analysis of IR spectroscopic
data, which provides new insights into the insulator-metal transition in polymers.

\begin{figure}
\includegraphics[width=9cm]{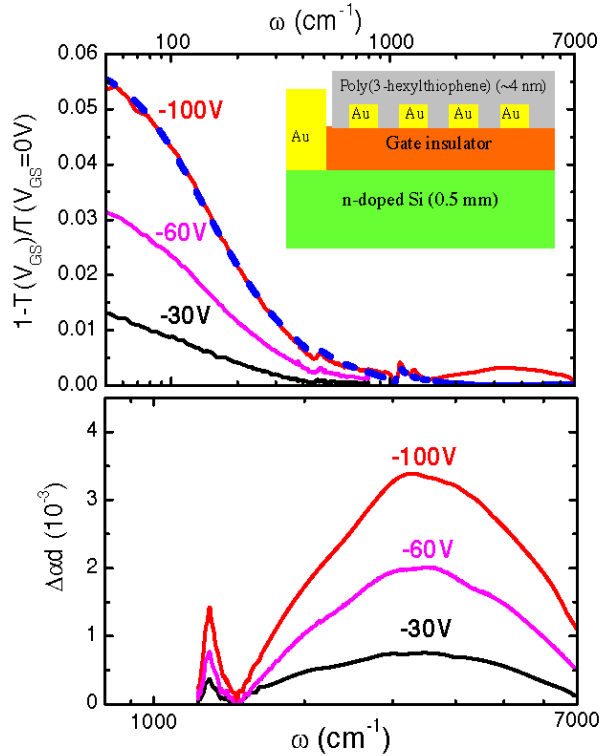}
\caption{(Color online) Top panel: the 1-T(V$_{GS}$)/T(0V) spectra for a representative
SiO$_{2}$ based organic FET device under applied gate voltages V$_{GS}$. Solid
curves: experimental spectra at several gate voltages. Blue dashed curve: the
spectrum at $-100$V obtained from fitting the low-energy absorption of the accumulation layer in n-Si
using the DL model as described in the text. Top inset: a schematic of the cross-section of the FET
devices. Bottom panel: The mid-IR voltage-induced absorption spectra $\Delta\alpha d$
for the P3HT layer, which are obtained by subtracting absorption due to the
accumulation layer in n-Si from the absorption spectra of the device as detailed
in the text.}
\label{fig:IR}
\end{figure}

Large area organic FET devices\cite{Li06p224,Li05p3506} ($>1$ cm$^{2}$) with gate insulator
deposited on n-Si were investigated in this work. In these devices, source and drain Au
electrodes (with a spacing of 50-200 $\mu$m) were patterned on a 200-nm-thick SiO$_2$ gate
insulator followed by the deposition of a 4--6 nm-thick P3HT film. The inset of Fig.~\ref
{fig:IR} displays a schematic of the cross-section of the device. These devices has transport
mobility of about 0.18 cm$^{2}$V$^{-1}$s$^{-1}$ and breakdown voltage in excess of
$-100$V.\cite{Wang03p6137} In these bottom-contact FET devices, an applied gate voltage induces
an accumulation layer\cite{Dimitrakopoulos02p99,Li02p4312} in P3HT as well as in n-Si. The
former accumulation layer constitutes the p-type conducting channel between the Au electrodes.
This channel is not obscured by any other interfaces and is therefore well suited for the
spectroscopic studies of the accumulation layer in the polymer film from far-IR to near-IR with
the latter cut-off imposed by the band gap of the Si substrate.\cite{Li06p224} We studied
changes of transmission as a function of applied gate voltage V$_{GS}$ normalized by the
transmission at V$_{GS}$=0: T($\omega $,V$_{GS}$)/T($\omega $,V$_{GS} $=0). Note
that in the same sample structure without a P3HT layer, we did not
observe any gate voltage induced changes in the IR spectrum. This
suggests that no accumulation layer forms in the region between the
gold grid electrodes without the P3HT layer on the top of the
insulator. The source and drain voltage is zero in most measurements. All the data reported here were recorded at room temperature with a spectral resolution of 4 cm$^{-1}$.

We start by analyzing the absorption spectra $\Delta \alpha $d = 1-T(V $_{GS}$)/T(0V) of the
organic FET devices displayed in Fig.~\ref{fig:IR}.  Here $\Delta \alpha $ is the change of the
absorption coefficient of P3HT and d is the thickness of the accumulation layer where absorption
is formed. These spectra show several voltage-induced features: i) a strong absorption in far-IR
frequencies ($\omega < 500$ cm$^{-1}$), ii) sharp resonances in the 1000 -- 1500 cm$^{-1}$
region, and iii) a broad band centered around 3500 cm$^{-1}$. A gradual development of these
features with increasing gate voltages V$_{GS}$ suggests that i-iii) are intimately related to
the formation of charge accumulation layers on both sides of the oxide.  Features ii) and iii)
are spectroscopic signatures of electrostatically doped carriers in P3HT.\cite{Li06p224} Sharp
resonances in the 1000 -- 1500 cm $^{-1}$ range result from the IR active vibrational modes
(IRAVs);\cite{Heeger88p781} i.e Raman modes made IR active by distortions of the polymer
backbone caused by the self-localized charges. The broad absorption band centered around 3500
cm$^{-1}$ is usually ascribed to polaron or bipolaron states positioned in the
gap\cite{Heeger88p781}; the lineshape of the broad band does not allow one to discriminate
between the two types of excitations. Both the IRAV modes and the polaron band can be
quantitatively described\cite{Li06p224} by the amplitude mode model\cite{Horovitz04p179} of
charge excitations in conjugated polymers. Features ii) or iii) are significantly weaker than
the far-IR absorption i). We attribute the latter to the charge accumulation layer on the n-Si
side of the gate dielectric. The validity of this assignment is supported by quantitative
agreement of the far-IR data with results for heavily doped Si\cite{Gaymann93p3681} as will be
described in detail below.
\begin{table}[tbp]
\caption{The parameters of the Drude-Lorentz fitting (Eq.~(\ref{eq:epsilon})) of the absorption
of the accumulation layer in n-Si of the FET device at V$_{GS}=-100$V. We use the ``effective
2D'' oscillator strengths $\protect\omega _{PD}^{2}$ and $\protect\omega _{P}^{2}$ associated
with 2D carrier densities N$_{f}$ and N$_{M}$, respectively.}
\begin{ruledtabular}
\begin{tabular}{cccc}
$\omega _{PD}^{2}$ & 2.57cm$^{-1}$ & N$_{M}$ & $0.28\times 10^{13}$cm$^{-2}
$ \\
N$_{f}$ & $0.74\times 10^{13}$cm$^{-2}$ &$\Gamma $ & 1400cm$^{-1}$  \\
$\Gamma _{D}$ & 132cm$^{-1}$ & $\omega _{0}$ &  360cm$^{-1}$\\
$\omega _{P}^{2}$ & 0.96cm$^{-1}$ & $\varepsilon _{\infty }$ & 10\\
\end{tabular}
\end{ruledtabular}
\label{tab:para}
\end{table}

We now turn to the analysis of the accumulation layer (AL) in n-Si.  These results will help us
to quantify the 2-dimensional (2D) carrier density which is the same on both sides of the gate
dielectric in the P3HT/insulator/n-Si structure. This task can be accomplished for example by
fitting the low-energy absorption with the Drude-Lorentz (DL) model, in which the complex
dielectric function $\varepsilon $($\omega$) has the form:
\begin{equation}
\varepsilon (\omega)=-\frac{\omega_{pD}^2}{\omega^2+i\Gamma_D\omega }+\frac{\omega_{p}^{2}}{\omega_{0}^{2}-\omega^{2}-i\Gamma\omega}+\varepsilon_{\infty}.
\label{eq:epsilon}
\end{equation}
The first term in Eq.~(\ref{eq:epsilon}) stands for the Drude free carrier response with the
relaxation rate $\Gamma_{D}$ and the plasma frequency $\omega_{PD}$ given by $\omega_{\rm
PD}^{2}=\frac{4\pi N_{\rm f}e^{2}}{m_{\rm b}}$, where N$_{\rm f}$ is the 2D density of free
carriers and $m_{\rm b}= 0.26 m_{\rm e}$ is the conduction band mass of Si. The second term is
used to account for small deviations of the dielectric function in mid-IR from the Drude form.
The last term $\varepsilon_\infty$ is the high frequency dielectric constant. Two scenarios are
the most likely origins of the second term: frequency dependence of the
relaxation rate of free carriers and/or interband transitions from donor impurities to the
conduction band of Si at $\omega_{0}$. The interband transitions are characterized by the
relaxation rate $\Gamma$ and the plasma frequency $\omega_P$ given by $\omega_P^{2}=\frac{4\pi
N_Me^2}{m^*}$, where $N_{\rm M}$ is the 2D charge density and $m^*\sim m_b$ is the renormalized
mass. It is advantageous to carry out the analysis using the ``effective 2D'' carrier density in
Eq.~(\ref{eq:epsilon}) in order to avoid any assumptions about the thickness of the accumulation
layer in n-Si.  The total 2D carrier density $N_{\rm 2D}=N_{\rm f}+N_{\rm M}$ represents a lower
limit of the density of the voltage-induced charges. 

In order to evaluate the consistency of the DL model with the data, we first obtained optical
constants for each of the constituting layers of our devices as described in
Ref.~\onlinecite{Li05p3506} and then considered the transmission of the multi--layered sample:
P3HT/SiO$_{2}$/2D-AL/n-Si. Results shown in the top panel of Fig.~\ref{fig:IR} suggest that the
experimental spectra can be adequately described by modeling the AL response using
Eq.~(\ref{eq:epsilon}) with parameters presented in Table ~\ref{tab:para}. Here we assumed that
voltage induced changes of P3HT in far-IR are negligible small in accord with dc
measurements.\cite{Wang03p6137} The parameters we obtained for the response of AL are in good
agreement with the optical spectra of metallic phosphorus doped Si.\cite{Gaymann93p3681} The
voltage-induced absorption spectra in the mid-IR frequencies for the P3HT layer can be obtained
by subtracting the absorption due to the accumulation layer in n-Si from the absorption spectra
of the device, which are displayed in the bottom panel of Fig.~\ref{fig:IR}.

\begin{figure}
\includegraphics[width=9cm]{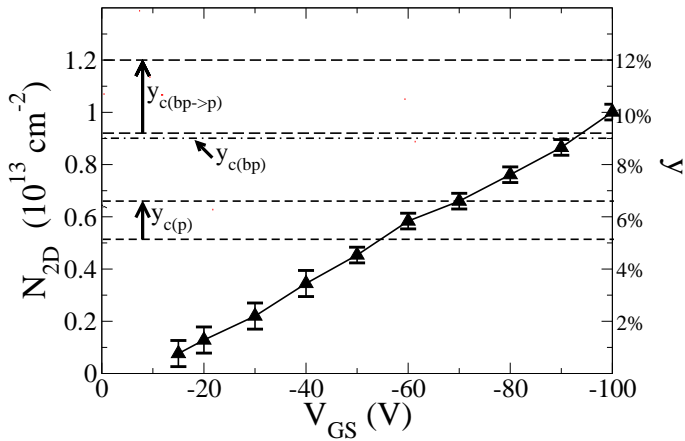}
\caption{The 2D carrier density of charges and the corresponding doping level $y$ (solid triangles, solid line is to guide the eye) induced in the P3HT polymer as a function of the gate voltage determined as described in the text. The density reaches $10^{13}$ cm$^{-2}$ (equivalent to 10\%) at the highest $V_{\rm GS}$ applied. The horizontal lines mark the different ranges of the critical dopant concentration at which metallic states
can be expected predicted from the theoretical models in Sec.~\ref{cross}--\ref{bip}.}
\label{fig:density}
\end{figure}

In Fig.~\ref{fig:density} we plot the 2D carrier density of charges $N_{\rm 2D}$ injected in
P3HT polymers obtained from the previous analysis as a function of the gate voltage $V_{\rm
GS}$. The carrier densities in the polymers reaches $10^{13}$ cm$^{-2}$ at the highest voltage.
Therefore, we have shown that the carrier density in P3HT in our organic FETs at the highest
gate biases is about $N_{\rm 3D}\sim 10^{20}$ cm$^{-3}$ corresponding to approximately one
charge in every 10 thiophene rings, i.e., $y\cong 10\%$.

\section{Discussion}
\label{dis}
\subsection{Comparison with the IR spectra}

Let us now compare the theoretical predictions with the IR spectroscopic results and discuss
their implications. In Sec.~\ref {energy}, we have estimated the lowest optical absorption
frequency to be 0.16eV for polarons and 0.46 eV for bipolarons. In our IR spectroscopic studies,
the localized absorption band due to charge injection is found at about 0.45 eV. The frequency
is appreciably higher than the predicted polaron absorption, while it matches reasonably well
with the predicted bipolaron absorption. The result seems to suggest that the localized band is
due to bipolaron excitations. On the other hand, polaronic energies reported in previous work
seem to vary appreciably suggesting a dependence on the sample qualities and experimental
platforms. Indeed, very broad range of polaron energies ranging from 0.06eV to 0.35eV has been
reported in different P3HT
systems.\cite{Sirringhaus99p685,Osterbacka02p226401,Wohlgenannt04p241204} Considering that the simple model followed in this paper assumes idealized 1D infinite chains, there is a need to
discuss in some details the disorder effect on the localized bands and the optical transition
energies. This discussion is presented in the subsequent subsection. 

The calculations in Sec.~\ref{theory} show that a relatively low doping concentration of about
5\% is expected to close the gap between the polaron bands and the continuum states, leading to
a possible second--order transition in P3HT. In our actual experiments, with a gate voltage of
$-100$V applied, the charge injection of the P3HT polymers achieves a doping concentration as
high as 10\% (see Fig.~\ref{fig:density}), exceeding the threshold over which the polaron--
related transition occurs. However, we do not have any direct evidence of metallic states, for
example, through reduction of the oscillator strength of the IRAV modes near this doping
range.\cite{Lee06p65} The insulating state at the doping level of 5\% is thus consistent with
our bipolaron interpretation of the spectra.

On the other hand, our results indicate that metallic states in P3HT can be expected at a doping
concentration of about 10\% due to either a first--order transition to the metallic polaron
lattice or a possible second--order transition by closing the gap between the bipolaron band and
the valence band. Both thresholds are close to the highest doping achieved in our
charge--injected polymers. In the following we suggest two issues that can lead to the absence
of direct evidence of metallic signatures in the present experiment.

First, recent experiments in P3HT FET suggested that at low temperature, a DC field is needed in
order to overcome the tunneling barriers between the occupied localized states which can
suppress conductivity of polymers and result in insulating properties.\cite{Dhoot06p246403}
Therefore while it is possible that the present P3HT system might have achieved a metallic
carrier density, it is not clear that one can observe metallic conductivity, as found recently
that only a small faction of the field-induced charge contributes to metallic conduction while
most charges occupying the localized states.\cite{Dhoot06p246403} Second, in the FET device of
interest here, the absorption in the accumulation layer on the n-Si side of the gate dielectric
dominates the observed spectrum in the far-IR frequencies, because of its much higher mobility
compared to that of P3HT. As a result, it is difficult to directly subtract the Drude absorption
due to the n-Si from the absorption of the device so that the Drude absorption and thus the
metallic signatures of the polymer at the far-IR frequency can be extracted.

\subsection{Effect of disorder}

In quasi-1D conducting polymers, disorder is expected to have a particularly important role
affecting both transport and other phenomena compared to systems with higher dimensions. Indeed,
previous work has suggested that insulator metallic transitions in {\it trans}--PA can result
from disorder in the polymers.\cite{Mele81p5397} Here we discuss the sensitivity of our
predictions to disorder of the polymers in terms of the confinement parameter $\gamma = \frac
{\Delta_e}{\lambda\Delta_0}$, which is directly related to the energies of the localized states.

In the presence of a higher degree of disorder, we expect that the degree of the conformational
nondegeneracy would be lowered due to the configurational randomness.  Hence the external gap
parameter $\Delta_e$ is lower compared to that of more regio-regular polymers. More importantly,
disorder affects the confinement parameters through the electron-phonon couplings, the strength
strongly influences the electronic structure of organic polymers.\cite{Phillpot87p7533} In
organic semiconductors, the electron-phonon coupling constant is related to the conformational
disorder through the degree of the chain conjugation. The coupling decreases with the
conjugation length,\cite{Shuai94p301} i.e., the length of the polymer segments separated by
defects. Hence, the electron-phonon coupling increases with disorder in the conjugated polymers.
On the basis of these arguments, one can anticipate that in the presence of disorder, the
confinement parameters $\gamma$ would be smaller than that in polymers with higher
regio-regularity.

In Sec.~\ref{cross}--\ref{bip}, we have carried out calculations of the critical densities in
P3HT using $\gamma\cong0.4$. Given the trend that $\gamma$ decreases with increasing disorder,
it is instructive to recalculate the results at values of $0< \gamma < 0.4$. To this end, we
have recalculated the values using $\gamma = 0.01$ and obtained $\omega_0 \cong 0.75$eV and
$\omega_0 \cong 0.1$eV that correspond to the lowest optical absorption energies of $0.3$eV and
$0.95$eV for the polarons and bipolarons, respectively. Note that the two bipolarons bands merge
into a single mid-gap state at $\gamma\cong0$.  The results show that both the polaron and the
bipolaron absorptions blueshift with increasing disorder of the polymers. However, even at the
smallest value of $\gamma$, it appears that the localized absorption in our experiment is still
considerably higher compared to the predicted polaron absorption, thus supporting our assignment
of the localized band as due to bipolarons, rather than polarons.

Extending this analysis to the larger end of $\gamma$, on the other hand, we find the lowest
polaron absorption energies reduced significantly to less than $0.1$eV when $\gamma\cong 1$.
This value is consistent with the recent photoinduced absorption measurement by \"Osterbacka
{\it et al.} of highly regio-regular P3HT polymers,\cite{Osterbacka02p226401} where the polaron
absorptions are found to superpose with the IRAV modes at very low frequencies less than 0.1eV.
Similar absorption features were also revealed by Brown {\it et al.} using charge modulation
spectroscopy in P3HT based MOS diode, albeit assignment of the polaron absorptions were given
somewhat differently.\cite{Brown01p125204} It is important to note that a distinction exists
between the RR P3HT in our experiment and those in the earlier
work,\cite{Osterbacka02p226401,Brown01p125204} despite both exhibit high mobilities. In our
experiment, the charge accumulation layer is restricted to the polymer/gate dielectric interface
within 1-2 nm of the 4-6 nm-thick P3HT films due to the bottom contact FET structure, the
interfacial disorder\cite{Kline06p222} would play a much more important role than in the
``bulk'' form of P3HT in the earlier experiments. We thus suggest that the difference between
the absorption peak we observed and those from the earlier work might be related to the extent
of the interfacial disorder present in the samples. 

We have also recalculated the doping thresholds at $0<\gamma < 0.4$.  This leads to a range of
the critical doping values as shown in Fig.~\ref {fig:density}. For the first two cases
discussed in Sec.~\ref{cross} and ~\ref{pol}, the thresholds increase slightly with decreasing
$\gamma$, i.e. with increasing disorder. For the bipolaron gap closure discussed in Section~\ref
{bip}, the value of $\gamma$ is lower bounded by the value at which the two bipolaron bands in
the gap touch due to the band broadening, which corresponds to $\gamma\cong0.3$. In other words,
below this value, the bipolaron gap closure condition can not be satisfied. In the range of
$0.3<\gamma<0.4$ , the critical doping concentration is nearly unchanged. We have indicated the
latter threshold by a single line in Fig.~\ref {fig:density}. Given the fact that these ranges
appear relatively insensitive to the structural disorder described by $\gamma$, together with
the assignment of the carries to bipolarons, we may conclude that the maximum carried densities
achieved in the present experiments are in close proximity to the regime where a first--order
transition from bipolaron lattice to metallic polaron lattice occurs.

We conclude this section by stressing that due to contact and interface effects, we expect the
precise value of the critical density for insulator-metal transition to be larger than the one
we have predicted above. In this respect, we compare our results to very recent experimental
data obtained by Panzer {\it et al.}\cite{Panzer06p1051} and Dhoot {\it et al.}\cite{Dhoot06-2}
The first authors have reported field-induced charge injection as high as $10^{22}$cm$^{-3}$ in
P3HT FETs with polyelectrolyte gate dielectric. At such densities -- higher than our predicted
minimal density -- P3HT indeed shows metallic transport properties. The second
group~\cite{Dhoot06-2} finds strong evidence of metallic behavior in a polymer which is
structurally very similar to P3HT. This metallic behavior is again observed at carrier densities
larger than the ones we predict. These results therefore corroborate our prediction that a
critical carrier density exists beyond which metal-insulator transition can be expected in P3HT.

\subsection{Limitation of the model}
We briefly comment on the limitations of our theoretical analysis.
The BK model used in this paper does not directly account for the
electron-electron interactions. In spite of ongoing efforts to
incorporate electron-electron interactions in the models of
conjugated polymers,\cite{Heeger88p781therein} it
is currently far from clear how that has to be done. Nonetheless, it
is worth noting that to some extent, electron-electron interactions
are effectively included in our calculations through the use of
experimental values for the parameters such as the band gap, the
characteristic correlation length, and the confinement parameters. We expect
that explicit inclusion of the interactions in the present treatment
may modify somewhat the critical concentrations predicted in the
analysis. On the other hand, the theoretically calculated
polaron/bipolaron energies and the experimental values obtained from
our P3HT FET are in reasonable agreement. This agreement and the
insensitivity of the results to the doping concentration, and hence
to the strength of the electron-electron interactions suggest that
the electron-lattice interactions are still the dominating effect in
P3HT polymers. The electron-phonon coupling model we use thus
captures the essential physics of localized excitations.

The BK model also does not account for the electric field. While the
effect of the electric field is important for understanding polaron
dynamics, we expect that it does not play an important role in
regioregular P3HT-based devices considered in this paper. In the
P3HT structures we have studied, the P3HT chains form lamellae
normal to the substrate with the chains oriented parallel to the
substrate surface.\cite{Sirringhaus99p685} As the applied gate field
is perpendicular to the film surface, we expect that the P3HT chains
are not subject to the gate field in the parallel direction and thus
the electrostatic potential along each chain is constant.

\section{Conclusion}
In this paper, we have analyzed the conditions to obtain metallic behavior in charge-injected
P3HT polymers through a first-order phase transition between bipolaron and polaron lattices or
gap closure between txshe bipolar/polaron bands and the continuum bands. IR spectroscopic
measurements in P3HT thin film FET show that the highest charge injection level achieved in the
present experiments is in proximity to the predicted doping thresholds for the first case,
indicating that metallic behavior can be anticipated. Optical absorption frequencies are
predicted for both polarons and bipolarons. The frequencies associated to bipolarons are in good
agreement with the absorption peak observed. Both conclusions remain true even when disorder is
taken into account. We therefore conclude that the localized excitations we observe are likely
to be due to bipolarons and a first-order transition from the bipolaron lattice to the polaron
lattice can be expected.

\begin{acknowledgements}
We acknowledge G.M. Wang, D. Moses and A.J. Heeger for providing the P3HT FET samples. The
research is supported by NSF grant NSF-0438018.
\end{acknowledgements}


\end{document}